\documentclass[aps,showpacs,tightenlines,twocolumn,nofootinbib,nobibnotes,superscriptaddress]{revtex4-1}
\usepackage{amsmath,amssymb,amsfonts,bm}
\usepackage{graphicx}
\usepackage{epstopdf}
\usepackage{dcolumn}
\usepackage{mathrsfs}
\usepackage{tgtermes}
\usepackage[colorlinks=true,linkcolor=blue,citecolor=blue, urlcolor=blue,bookmarks=false]{hyperref}
\usepackage{lipsum}
\bibpunct{[}{]}{,}{n}{}{}
\usepackage{multirow}
\begin{document}
\title{Higher-order topological Anderson insulators in quasicrystals}
\date{\today }
\author{Tan Peng}
\affiliation{Department of Physics, Hubei University, Wuhan 430062, China}
\author{Chun-Bo Hua}
\affiliation{Department of Physics, Hubei University, Wuhan 430062, China}
\author{Rui Chen}
\affiliation{Shenzhen Institute for Quantum Science and Engineering and Department of Physics, Southern University of Science and Technology (SUSTech), Shenzhen 518055, China}
\author{Zheng-Rong Liu}
\affiliation{Department of Physics, Hubei University, Wuhan 430062, China}
\author{Dong-Hui Xu}
\affiliation{Department of Physics, Hubei University, Wuhan 430062, China}
\author{Bin Zhou}\email{binzhou@hubu.edu.cn}
\affiliation{Department of Physics, Hubei University, Wuhan 430062, China}

\begin{abstract}
The disorder effects on higher-order topological phases in periodic systems have attracted much attention. However, in aperiodic systems, such as quasicrystalline systems, the interplay between disorder and higher-order topology is still unclear. In this paper, we investigate the effects of disorder on two types of second-order topological insulators, including a quasicrystalline quadrupole insulator and a modified quantum spin Hall insulator, in a two-dimensional Amman-Beenker tiling quasicrystalline lattice. We demonstrate that the higher-order topological insulators are robust against weak disorder in both models. More striking, the disorder-induced higher-order topological insulators called higher-order topological Anderson insulators are found at a certain region of disorder strength in both models. Our paper extends the study of the interplay between disorder and higher-order topology to quasicrystalline systems.

\end{abstract}

\maketitle

\section{Introduction}
A higher-order topological insulator (HOTI), a generalization of conventional topological insulator (TI), has been a hot point of research in condensed-matter physics \cite{saha2021higher,schindler2021tutorial,xie2021higher,Benalcazar61,PhysRevB.96.245115,Schindlereaat0346,PhysRevLett.119.246401,PhysRevLett.119.246402}. Unlike the conventional TI, an $n$th-order topological insulator that has $d$ dimensions will have gapless boundary states in $d-n$ dimensions ($d\ge n$). For instance, a two-dimensional (2D) second-order topological insulator (SOTI) has zero-dimensional (0D) corner states localized at its boundary. Analogously, a three-dimensional (3D) second- (third-) order topological insulator has one-dimensional (1D) (0D) hinge (corner) states localized at its boundary. These novel bulk-boundary correspondences, which are quite different from the conventional TIs, can be described by the nested-Wilson-loop method \cite{Benalcazar61,PhysRevB.96.245115,shang2020secondorder} and the real-space quadrupole moment \cite{Benalcazar61,PhysRevLett.119.246401,PhysRevLett.119.246402,Schindlereaat0346,PhysRevB.97.205135,PhysRevB.96.245115}.

The HOTIs have been extensively studied in various systems \cite{PhysRevB.97.205135,
Fangeaat2374,xu2017topological,PhysRevLett.120.026801,PhysRevLett.121.116801,PhysRevB.97.155305,PhysRevB.97.205136,PhysRevB.97.241402,
PhysRevB.97.241405,PhysRevB.98.045125,PhysRevLett.121.096803,PhysRevLett.121.186801,PhysRevB.97.094508,PhysRevB.98.235102,
PhysRevB.98.241103,PhysRevB.98.245102,PhysRevLett.123.016805,PhysRevLett.122.076801,PhysRevLett.122.204301,PhysRevLett.123.247401,
PhysRevLett.123.256402,PhysRevB.99.245151,PhysRevB.100.085138,PhysRevB.100.235302,PhysRevB.101.041404,PhysRevLett.122.236401,
PhysRevB.99.125149,PhysRevB.100.205406,PhysRevLett.125.097001,
PhysRevB.98.081110,PhysRevB.98.165144,PhysRevB.98.201114,PhysRevLett.124.216601,PhysRevX.9.011012,PhysRevLett.122.086804,
PhysRevLett.123.156801,PhysRevLett.123.167001,PhysRevLett.123.177001,PhysRevLett.123.186401,PhysRevLett.123.216803,PhysRevB.99.041301,
PhysRevB.99.235132,PhysRevResearch.3.013239,PhysRevB.102.094503,wieder2020strong,PhysRevResearch.2.033029,schindler2018higher,
serra2018observation,xue2019acoustic,ni2019observation,peterson2018quantized,mittal2019photonic,zhang2020low,noh2018topological,imhof2018topolectrical,
PhysRevB.100.201406}.
Up to now, the great majority of the previous works about the HOTIs were discussed in crystalline systems. However, the aperiodic systems, especially the quasicrystalline systems, which lack translational symmetry and possess forbidden symmetries in crystals, such as the fivefold, eightfold, and twelvefold rotational symmetries, have also been used to realize HOTIs  \cite{PhysRevLett.123.196401,PhysRevLett.124.036803,PhysRevB.102.241102,PhysRevResearch.2.033071}. For instance, Chen \emph{et al}. proposed that two distinct types of SOTIs can be realized in quasicrystalline lattices \cite{PhysRevLett.124.036803}. One is the quasicrystalline quadrupole insulator which can be constructed in a modified Benlcazar-Bernevig-Hughes model \cite{Benalcazar61,PhysRevB.96.245115}, and this kind of SOTI is protected by chiral symmetry. The other is the modified quantum spin Hall insulator which is formed by a TI model with a mass term which gaps the counterpropagating edge states and induces the appearance of topological corner states. They proved that these types of the topological corner states are protected by combined symmetries $C_{4}m_{z}$ and $C_{8}m_{z}$ with different boundary conditions. Very recently, Lv \emph{et al}. reported that the HOTI has been experimentally implemented in a quasicrystalline lattice constructed by electrical circuits \cite{lv2021realization}.

\begin{figure*}[tp]
    \centering
	\includegraphics[width=16cm]{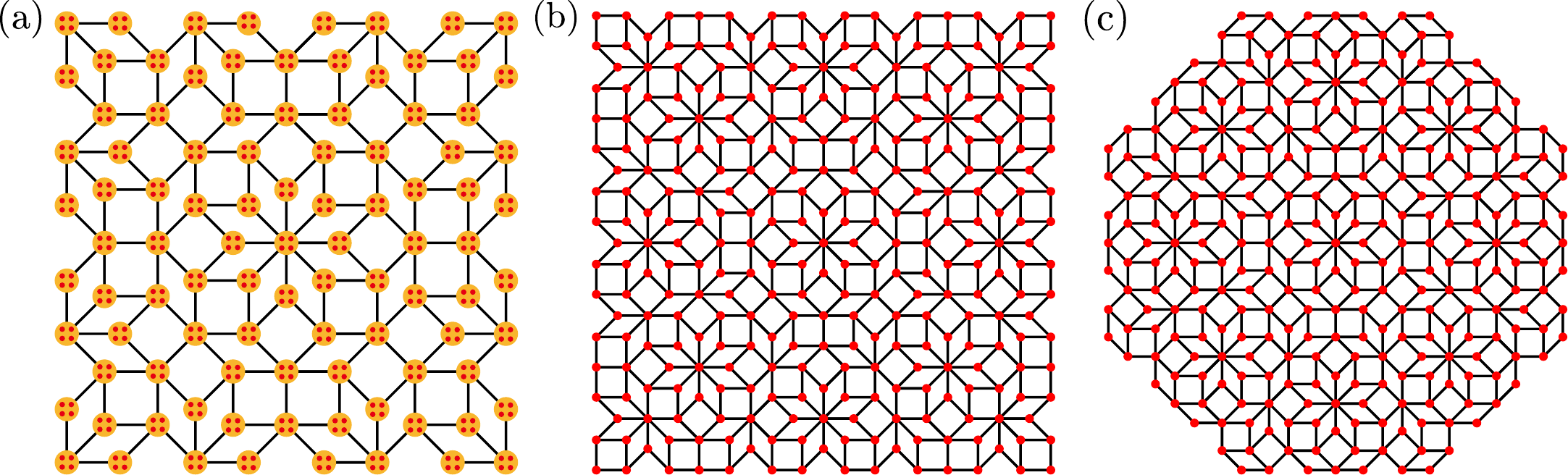} \caption{(a) Schematic of the Ammann-Beenker tiling quasicrystal containing $94$ cells. Each cell includes four sites, marked by orange. (b) An Ammann-Beenker tiling quasicrystal containing $301$ vertices with the square boundary condition. (c) An Ammann-Beenker tiling quasicrystal containing $297$ vertices with the octagonal boundary condition. The first three nearest-neighbor intercell bonds correspond to the short diagonal of the rhombus tile, the edge of square and rhombus tiles, and the diagonal of the square tile, respectively. The distance ratio of the three bonds is $r_{0}:r_{1}:r_{2}=2\sin \frac{\pi }{8}:1:2\sin \frac{\pi }{4}$.}%
\label{fig1}
\end{figure*}

Another interesting topic is disorder-induced topological phase transition. Generally, the topological phase is robust against weak disorder and suppressed by strong disorder where the energy gap is closed and a topological phase transition appears. Furthermore, a fascinating phenomenon is that disorder can encourage the generation of a topological phase by adding a certain strength of disorder to a topologically trivial phase. The disorder-induced topological phase which is a so-called topological Anderson insulator (TAI) was first proposed by Li \emph{et al}. in 2009 \cite{PhysRevLett.102.136806}. Then, the TAIs have been extensive studied in various systems \cite{PhysRevLett.105.115501,Zhang_2013,PhysRevB.92.085410,PhysRevLett.116.066401,PhysRevB.100.054108,PhysRevB.80.165316,
PhysRevLett.103.196805,PhysRevB.83.045114,PhysRevB.91.214202,PhysRevB.96.205304,PhysRevB.84.035110,orth2016topological,
PhysRevLett.105.216601,PhysRevB.82.115122,PhysRevLett.113.046802,PhysRevLett.115.246603,PhysRevB.93.075108,PhysRevB.95.245305,
PhysRevB.97.235109,PhysRevB.98.235159,PhysRevB.97.024204,PhysRevB.93.125133,qin2016disorder,PhysRevB.98.134507,PhysRevB.100.205302,
PhysRevA.101.063612} and realized in several experiment platforms \cite{meier2018observation,stutzer2018photonic,PhysRevLett.125.133603}, such as the 1D-disordered atomic wires \cite{meier2018observation} and the photonic lattices \cite{stutzer2018photonic,PhysRevLett.125.133603}. TAI was implemented in the crystalline systems, as well as in quasicrystalline systems. For instance, TAI has been proposed in the Penrose tiling pentagonal quasicrystal \cite{PhysRevB.100.115311} and the Ammann-Beenker tiling octagonal quasicrystal \cite{PhysRevB.103.085307,hua2021disorderinduced}. In addition, the disorder-induced HOTI, dubbed the higher-order topological Anderson insulator (HOTAI), has been studied in various condensed-matter systems, including topological quadrupole insulators \cite{PhysRevLett.125.166801,PhysRevB.103.085408}, topological superconductors \cite{PhysRevB.100.075415}, and topological Weyl semimetals \cite{zhang2021global}. Even more striking is that Zhang \emph{et al}. demonstrated that the HOTAI can appear in a modified Haldane model and realize experimentally in the electric circuit setup \cite{PhysRevLett.126.146802}. However, the investigation of the disorder-induced HOTI in quasicrystalline systems remains lacking. An interesting question is whether the HOTAI can be realized in quasicrystalline systems.

In this paper, we investigate the disorder-induced topological phase transition in an Ammann-Beenker tiling octagonal quasicrystal with two types of SOTI containing a quasicrystalline quadrupole insulator (named model I) and a modified quantum spin Hall insulator with gapless corner states (named model II) as mentioned in Ref. \cite{PhysRevLett.124.036803}. For model I, the lattice is cut into a square, and each cell contains four sites as shown in Fig.~\ref{fig1}(a). For model II, we divide the discussion into two cases: a square boundary [shown in Fig.~\ref{fig1}(b)] and an octagonal boundary [shown in Fig.~\ref{fig1}(c)].  By calculating the quadrupole moment and the probability density of the in-gap eigenstates, we find that the SOTI phases in the two models with the square boundary conditions are robust against the weak disorder, and the HOTAI phase induced from an initial topological trivial phase occurs at a certain area of disorder strength with four localized gapless corner states characterized by a quantized quadrupole moment ($q_{xy}=0.5$). More striking, a HOTAI phase with eight localized gapless corner states is found in model II with an octagonal boundary, and this HOTAI phase is a unique topological phase which cannot be realized in crystalline systems.

The rest of the paper is organized as follows. In Sec.~\ref{Models}, we introduce two types of SOTIs with disorder in the 2D quasicrystalline lattice and give the details of numerical methods. Then, we provide numerical results for studying the topological phase transitions of the two models in Secs.~\ref{Model1} and~\ref{Model2}, respectively. Finally, we summarize our conclusions in Sec.~\ref{Conclusion}.

\section{Models and Method}
\label{Models}
We start with a tight-binding model of a quadrupole insulator with disorder in an Ammann-Beenker tiling quasicrystalline lattice which has a square boundary condition. Each vertex of the quasicrystalline lattice contains four sites as shown in Fig.~\ref{fig1}(a). In this section, we consider the first three nearest-neighbor intercell hoppings and the nearest-neighbor intracell hopping. The model Hamiltonian is given by \cite{PhysRevLett.124.036803}
\begin{eqnarray}
H_{1} &=&\lambda \sum_{m\neq n}\frac{l(r_{mn})}{2}c_{m}^{\dagger
}(\left\vert \cos \psi _{mn}\right\vert \Gamma _{4}-i\cos \psi _{mn}\Gamma
_{3} \notag \\
&&+\left\vert \sin \psi _{mn}\right\vert \Gamma _{2}-i\sin \psi _{mn}\Gamma
_{1})c_{n}\notag \\
&&+\sum_{m}c_{m}^{\dagger }[(\gamma _{x}+U_{m})\Gamma _{4}+\gamma _{y}\Gamma
_{2}]c_{m},
 \label{H1}
\end{eqnarray}
where  $c_{m}^{\dag }=(c_{m1 }^{\dag },c_{m2}^{\dag
},c_{m3 }^{\dag },c_{m4 }^{\dag })$  is the creation  operator in cell $m$. $\gamma_{x,y}$ are the intracell hopping amplitudes along the $x$ axis and $y$ axis, respectively. $\lambda$ denotes the intercell hopping amplitude. $U_{m}$ is the uniform random variable chosen from $[-W/2, W/2]$, and $W$ is the disorder strength. $\Gamma_{4}=\tau_{1}\tau_{0}$ and $\Gamma_{\mu}=-\tau_{2}\tau_{\mu}$ with $\mu=1,2,3$. $\tau_{1-3}$ are the Pauli matrices acting on the sites in one cell, and $\tau_{0}$ is the $2\times 2$ identity matrix. In polar coordinate space, $\psi _{mn}$ is the polar angle between cells $m$ and $n$. $l(r_{mn})=e^{1-r_{mn}/\zeta }$ is the  spatial decay factor of hopping amplitudes with the decay length $\zeta $ and $r_{mn}$ representing the spatial distance of arbitrary two cells. Here, the spatial decay length $\zeta$ and the side length of the rhombus and square $r_{1}$ are fixed as $1$, and the energy unit is set as $\lambda=1$ for simplicity. The Hamiltonian $H_{1}$ respects time-reversal symmetry, particle-hole symmetry, and chiral symmetry in the clean limit, i.e., $W=0$, and the time-reversal symmetry, particle-hole symmetry, and chiral symmetry operators are $T=\tau_{0}\tau_{0} K$, $P=\tau_{3}\tau_{0} K$ and $S=TP=\tau_{3}\tau_{0}$, respectively, where $K$ is the complex conjugate operator. When the disorder strength is not zero, the system also maintains these three symmetries. In fact, the Hamiltonian $H_{1}$ is a derivation of the Benalcazar-Bernevig-Hughes model \cite{Benalcazar61,PhysRevB.96.245115} in some sense. In addition, we also will investigate the effects of disorder on the higher-order topological phase of a modified Bernevig-Hughes-Zhang model in the Ammann-Beenker tiling quasicrystalline lattice with the square boundary condition and octagon boundary condition, respectively. As shown in Figs.~\ref{fig1}(b) and ~\ref{fig1}(c), each vertex of the quasicrystalline lattice is exactly one lattice site. Only the first three nearest-neighbor hoppings are considered in our computing. This model lattice can be described by a tight-binding Hamiltonian with the form of  \cite{PhysRevLett.124.036803}
\begin{eqnarray}
H_{2} &=&-\sum_{m\neq n}\frac{l(r_{mn})}{2}c_{m}^{\dagger }[it_{1}(s_{3}\tau
_{1}\cos \psi _{mn}+s_{0}\tau _{2}\sin \psi _{mn}) \notag \\
&&+t_{2}s_{0}\tau _{3}+t_{3}s_{1}\tau _{1}\cos (\xi \psi _{mn})]c_{n}\notag \\
&&+\sum_{m}(M+2t_{2}+U_{m})c_{m}^{\dagger }s_{0}\tau _{3}c_{m},
\label{H2}
\end{eqnarray}

\begin{table*}[htpb]
\caption{Symmetries of the Hamiltonian $H_{2}$ on an Ammann-Beenker tiling quasicrystalline lattice with the square and octagonal boundaries without ($W=0$) and with ($W\neq0$) disorder. $\sigma_{x,y,z}$ and $\tau_{x,y,z}$ are the Pauli matrices. $K$ is the complex conjugate operator, and $\mathcal{I}$ is the $N\times N$ unit matrix with the lattice number $N$. $\mathcal{M}_{x,y,}$ are orthogonal matrices permuting the sites of the tiling to flip the whole system vertically and horizontally. $\mathcal{R}_{4,8}$ are orthogonal matrix permuting the sites of the tiling to rotate the whole system by an angle of $\pi/2$ and $\pi/4$, respectively. Check mark indicates that the symmetry in this case is preserved, and a cross mark means the symmetry is absent.}
    \label{tab1}
\centering
\begin{tabular}{p{4cm}<{\centering}|p{5cm}<{\centering}|p{2cm}<{\centering}|p{2cm}<{\centering}|p{2cm}<{\centering}|p{2cm}<{\centering}}
\hline\hline
\multirow{2}{*}{}&\multirow{2}{*}{}&\multicolumn{2}{c|}{Square}&\multicolumn{2}{c}{Octagon}\cr
\hline
&&$W=0$&$W\not=0$&$W=0$&$W\not=0$\cr
\hline
$P=\sigma_{z}\tau_{x}\mathcal{I}K$  &$PHP^{-1}=-H$                     &\checkmark &$\times$&\checkmark &$\times$              \\
\hline
$T=i \sigma_{y}\tau_{0}\mathcal{I}K$                        &$THT^{-1}=H$&$\times$&$\times$&$\times$&$\times$        \\
\hline
$S=PT$                        &$SHS^{-1}=-H$ &$\times$&$\times$&$\times$&$\times$            \\
\hline
$m_{x}=\sigma_{x}\tau_{0}\mathcal{M}_{x}$&$m_{x}Hm_{x}^{-1}=H$&\checkmark &$\times$&\checkmark&$\times$       \\
\hline
$m_{y}=\sigma_{y}\tau_{z}\mathcal{M}_{y}$&$m_{y}Hm_{y}^{-1}=H$&\checkmark &$\times$&\checkmark&$\times$          \\
\hline
$m_{z}= \sigma_{z}\tau_{0}\mathcal{I}$&$m_{z}Hm_{z}^{-1}=H$&$\times$ &$\times$&$\times$&$\times$    \\
\hline
& & & \\[-9pt]
$C_{4}=e^{-i\frac{\pi}{4}\sigma_{z}\tau_{z}}\mathcal{R}_{4}$&$C_{4}HC_{4}^{-1}=H$&$\times$&$\times$&$\times$&$\times$            \\
\hline
$C_{4}T$&$C_{4}TH(C_{4}T)^{-1}=H$&\checkmark      &$\times$    &$\times$&$\times$          \\
\hline
$C_{4}m_{x}$&$C_{4}m_{x}H(C_{4}m_{x})^{-1}=H$&$\times$      &$\times$ &\checkmark&$\times$             \\
\hline
$C_{4}m_{y}$&$C_{4}m_{y}H(C_{4}m_{y})^{-1}=H$&$\times$     &$\times$  &\checkmark&$\times$              \\
\hline
$C_{4}m_{z}$&$C_{4}m_{z}H(C_{4}m_{z})^{-1}=H$&\checkmark     &$\times$  &$\times$ &$\times$               \\
\hline
& & & \\[-9pt]
$C_{8}=e^{-i\frac{\pi}{8}\sigma_{z}\tau_{z}}\mathcal{R}_{8}$&$C_{8}HC_{8}^{-1}=H$&$\times$&$\times$&$\times$&$\times$            \\
\hline
$C_{8}T$&$C_{8}TH(C_{8}T)^{-1}=H$     &$\times$&$\times$&\checkmark &$\times$              \\
\hline
$C_{8}m_{x}$&$C_{8}m_{x}H(C_{8}m_{x})^{-1}=H$&$\times$&$\times$&$\times$  &$\times$               \\
\hline
$C_{8}m_{y}$&$C_{8}m_{y}H(C_{8}m_{y})^{-1}=H$&$\times$&$\times$&$\times$&$\times$              \\
\hline
$C_{8}m_{z}$&$C_{8}m_{z}H(C_{8}m_{z})^{-1}=H$&$\times$&$\times$&\checkmark&$\times$                \\
\hline\hline
\end{tabular}

\end{table*}

where $c_{m}^{\dag }=(c_{m\alpha \uparrow }^{\dag },c_{m\alpha \downarrow }^{\dag
},c_{m\beta \uparrow }^{\dag },c_{m\beta \downarrow }^{\dag })$ represents the creation operator of an electron on a site $m$. In each site, $\alpha$ ($\beta$) is the index of orbitals, and $\uparrow$ ($\downarrow$) represents the spin direction. $s_{1-3}$ and $\tau_{1-3}$ are the Pauli matrices acting on the spin and orbital degree of freedom, respectively. $s_{0}$ and $\tau_{0}$ are the $2\times 2$ identity matrices. $t_{1-3}$ are the hopping strength, and $M$ is the Dirac mass.  The term containing $t_{3}$ is actually equivalent to a mass term that destroys the time-reversal symmetry of the system so that the original helical boundary state of the system opens the energy gap and evolves into a higher-order corner state \cite{PhysRevLett.124.036803,PhysRevB.102.241102}. $\xi$ is the varying period of the mass term, and $\xi = 2$ (4) for square (octagonal) samples. In the clean limit, the Hamiltonian $H_{2}$ respects particle-hole symmetry, mirror symmetry $m_{x,y}$, and some combined symmetries, such as $C_{4}T$, $C_{4}m_{z}$ with the square boundary condition and $C_{4}m_{x}$, $C_{4}m_{y}$, $C_{8}T$, $C_{8}m_{z}$ with the octagonal boundary condition as shown in Table ~\ref{tab1}. In fact, it has been demonstrated that the higher-order corner state is protected by the combined symmetry $C_{4}m_{z}$ ($C_{8}m_{z}$) with the square (octagonal) boundary condition by employing some uniform perturbations to test the stability of the corner states \cite{PhysRevLett.124.036803}. However, all symmetries are broken when the disorder is introduced and whether the higher-order corner states induced by the disorder can appear is unclear. Without loss of generality, we will set $t_{1}=t_{2}=1$.

The nested-Wilson-loop method \cite{Benalcazar61,shang2020secondorder,PhysRevB.96.245115} in the momentum space is efficient to characterize the topological phase of an electric quadrupole insulator. However, the quasicrystalline lattice is the lack of the translation invariance, thus, the topological invariant defined in the momentum space is no longer applicable for our models of quasicrystals. Therefore, we employ a real-space quadrupole moment to characterize the topological phases of the quasicrystalline lattice with disorder. The real-space quadrupole moment is given by \cite{PhysRevB.100.245134,PhysRevB.100.245135,PhysRevB.101.195309,PhysRevLett.125.166801,PhysRevResearch.2.012067}
\begin{equation}
q_{xy}=\frac{1}{2\pi }{\rm{Im}} \ln [\det (\Psi _{occ}^{\dagger }\hat{U}\Psi
_{occ})\sqrt{\det (\hat{U}^{\dagger })}],
 \label{qxy}
\end{equation}
where $\Psi _{occ}$ is the eigenvector of occupied states. $\hat{U}\equiv \exp [i2\pi \hat{X}\hat{Y}/N]$ where $\hat{X}$ and $\hat{Y}$ are the position operators, and $N$ represents the total number of the lattice sites. If $q_{xy}=0.5$, the system is a SOTI phase with topological corner states. Besides, $q_{xy}=0$ indicates a trivial phase. Note that the subsequent calculations of $q_{xy}$ in this paper are based on the periodic boundary condition. It is also noted that the validity of the formulation of the bulk quadrupole moment proposed by  two previous works \cite{PhysRevB.100.245134,PhysRevB.100.245135} is still controversial. Ono \emph{et al}. presented that the proposed definition of the bulk quadrupole moment fails even for a simple noninteracting example \cite{PhysRevB.100.245133}. Thus, a satisfactory formulation of the bulk quadrupole moment should be worthy of further study in the future works.

\section{Model I: Chiral symmetry-protected higher-order topological insulator}
\label{Model1}
In this section, we focus on the disorder-induced topological phase transition with chiral symmetry in an Ammann-Beenker tiling quasicrystal with the square boundary condition. The disorder is of the $U_{m}\Gamma _{4}$ type, which does not destroy the chiral symmetry.

\begin{figure}[tp]
	\includegraphics[width=8.5cm]{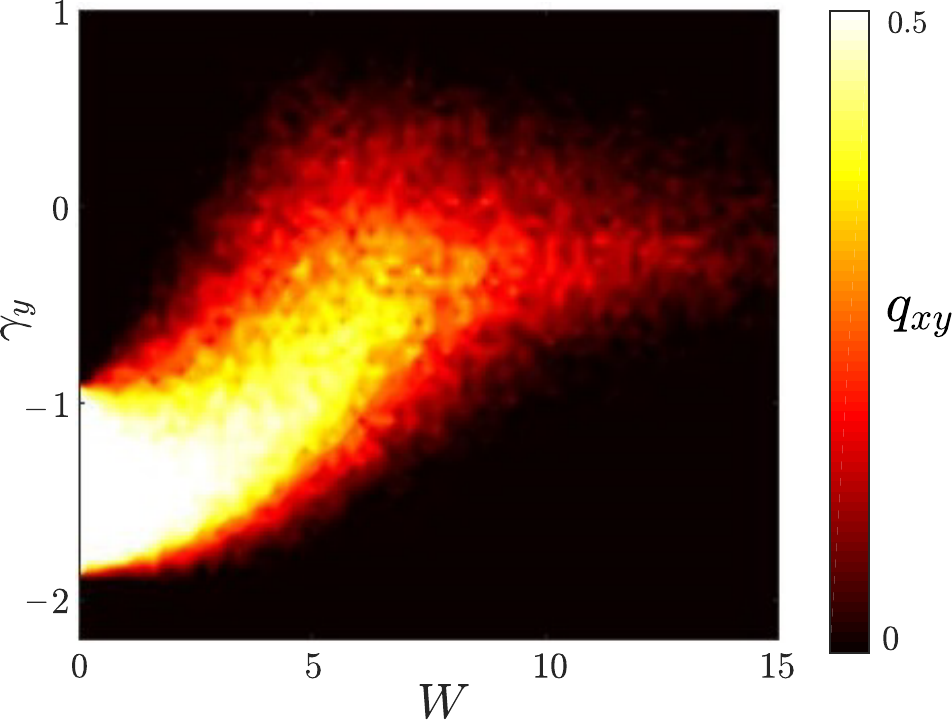} \caption{Topological phase diagram of the Ammann-Beenker tiling quasicrystal in ($W, \gamma_{y}$) space obtained by calculating the real-space topological invariant quadrupole moment $q_{xy}$ with $\gamma_{x}=-1.5$. The system is cut to a square sample containing $1257$ cells with periodical boundary conditions. Some $100$ random configuration averages are taken in our computing. }%
\label{fig2}
\end{figure}

Figure~\ref{fig2} shows the real-space quadrupole moment as a function of disorder strength $W$ and intracell hopping amplitude along the $y$-axis $t_{y}$ with fixed $t_{x}$. The color map shows the magnitude of the real-space quadrupole moment. It is found that when $W=0$, that is, in the clean limit, the system is in a SOTI phase with $q_{xy}=0.5$ if $\gamma_{y}$ satisfies $-1.8<\gamma_{y}<-0.9$. However, with the gradual increase in the disorder strength, the SOTI phase will transform to the Anderson localized state phase with $q_{xy}$ changing from $0.5$ to $0$. There are a series of critical maximum disorder strengths increased monotonically with the increase in $\gamma_{y}$. Meanwhile, we also find a phase which is a disorder-induced SOTI phase in the region where $\gamma_{y}>-0.9$. However, in our calculation, $q_{xy}$ is not a quantum number strictly equal to $0.5$ in the phase region. We believe that this is due to the finite-size effect of the system and will be discussed in the follow-up.

\begin{figure}[tp]
	\includegraphics[width=9cm]{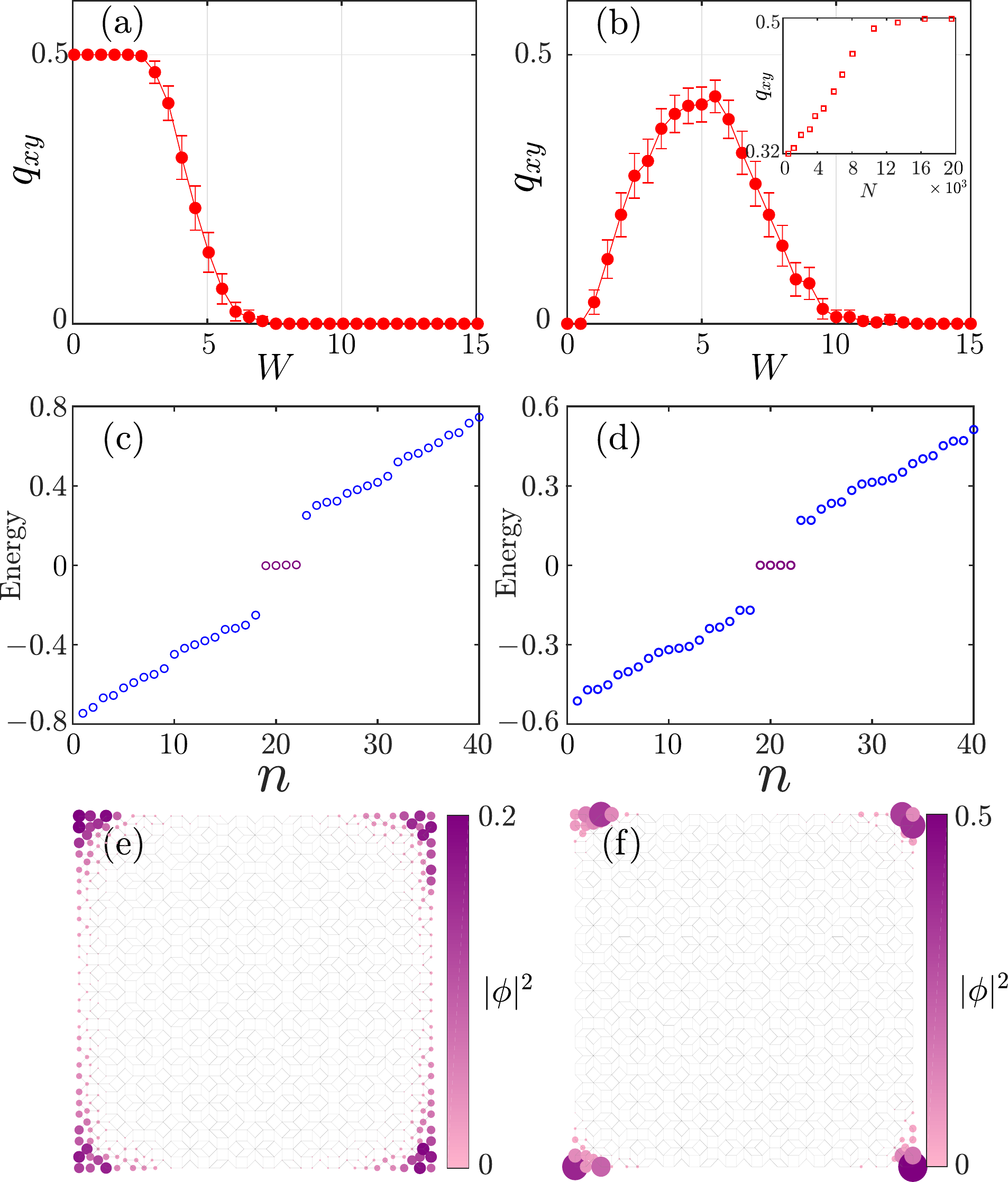} \caption{The real-space quadrupole moment $q_{xy}$ versus disorder strength $W$ with different initial states including (a) a higher-order topological phase with $\gamma_{x}=-1.5, \gamma_{y}=-1.5$ and (b) a topological trivial phase with $\gamma_{x}=-1.5, \gamma_{y}=-0.75$. The periodic boundary condition is taken, and 500 disorder configurations are performed. The inset shows the quadrupole moments $q_{xy}$ versus $N$ when $W=5.5$. $N$ is the total number of the cells. The energy modes near the zero energy for (c) a higher-order topological initial phase with $\gamma_{y}=-1.5$, $W=1.5$ and (d) a trivial initial phase with $\gamma_{y}=-0.75$, $W=4$, respectively. (e) and (f) The wave-function distribution of the zero modes corresponds to (c) and (d), respectively. The system contains $1257$ cells, and the open boundary condition is taken in (c)-(f).} %
\label{fig3}
\end{figure}

In order to explore the role of the disorder effect in the quasicrytalline lattice with chiral symmetry in more depth, we take two specific parameter values of $\gamma_{y}$ and plot the variation of $q_{xy}$ with respect to the strength of disorder as shown in Figs.~\ref{fig3}(a) and \ref{fig3}(b). For the case of $\gamma_{y}=-1.5$, the second-order phase remains stable in a weakly disordered situation ($W<2.5$) with a quantized quadrupole moment plateau and is destroyed in the strongly disordered situation ($W>7$) where $q_{xy}=0$. On the other hand, when $\gamma_{y}=-0.75$, the system hosts a trivial phase with $q_{xy}=0$ in the clean limit. As the strength of disorder increasing, $q_{xy}$ gradually increases from $0$ and approaches $0.5$, indicating that the system has undergone a phase transition from a trivial phase to a topological nontrivial phase. Actually, there is not a quantized quadrupole moment plateau in Fig.~\ref{fig3}(b), and we attribute this to the finite-size effect. Therefore, we plot $q_{xy}$ versus system size $N$ when $W=5.5$ with $500$ disorder configurations in the inset in Fig.~\ref{fig3}(b). It is found that $q_{xy}$ approaches $0.5$ with a large system size ($N=16437$). To further certify the existence of SOTI phases, we set some specific values of $W$ in Figs.~\ref{fig3}(a) and \ref{fig3}(b) to give the energy spectrum and wave-function diagram of the system. It is found that the SOTI phase is robust against the weak disorder [$W=1.5$ in Figs.~\ref{fig3}(c) and \ref{fig3}(e)] since the system hosts four zero-energy modes which are localized at the four corners of the lattice. This property is similar to the first-order topological state. Similarly, when $W=4$, four zero-energy modes appear at the four corners of the lattice which indicate the presence of the disorder-induced SOTI phases. The corner states are protected by the chiral symmetry which is quite similar to the corner states that appeared in crystalline systems in some previous works \cite{PhysRevLett.125.166801,PhysRevB.103.085408}.

\section{Model II: Combined symmetry-protected higher-order topological insulator}
\label{Model2}

In this section, we concentrate on the effects of disorder on the combined symmetry-protected higher-order topological phase in an Ammann-Beenker tiling quasicrystal with square and octagonal boundary conditions, respectively. All calculations are based on the Hamiltonian $H_{2}$. In the clean limit, the HOTI phase is protected by combined symmetry $C_{4}m_{z}$ and $C_{8}m_{z}$ for different boundary conditions. So far, it was revealed that the HOTI phase is protected by the symmetries, such as chiral, partial-hole, $C_{4}T$, $C_{4}m_{z}$ and $C_{8}m_{z}$ symmetries. According to the previous work \cite{PhysRevLett.125.166801}, the values of $q_{xy}$ can be quantized to $0$ or $1/2$ only if the system has chiral or partial-hole symmetry. However, all of these symmetries are destroyed when the disorder is introduced in the Hamiltonian $H_{2}$ (see Table I). Hence, the real-space quadrupole moment discussed in Sec.~\ref{Model1} may not be appropriate for model II with the square boundary condition. In addition, there is no well-defined topological invariant for a lattice with an octagonal boundary condition. One appropriate way to characterize the higher-order topological phase is to adopt the existence of the corner states as a working definition \cite{PhysRevB.99.085406,PhysRevLett.126.146802}. Thus, we calculate the energy spectrum and wave-function distribution of the system to determine whether the corner states exist. To reveal HOTAI in the quasicrystal described by model II, in the following calculations, we will not only perform disorder configuration average by many enough times, but also try to ensure that the size of the samples is large enough.

\subsection{Square boundary condition}
\label{square}

\begin{figure}[tp]
	\includegraphics[width=8.5cm]{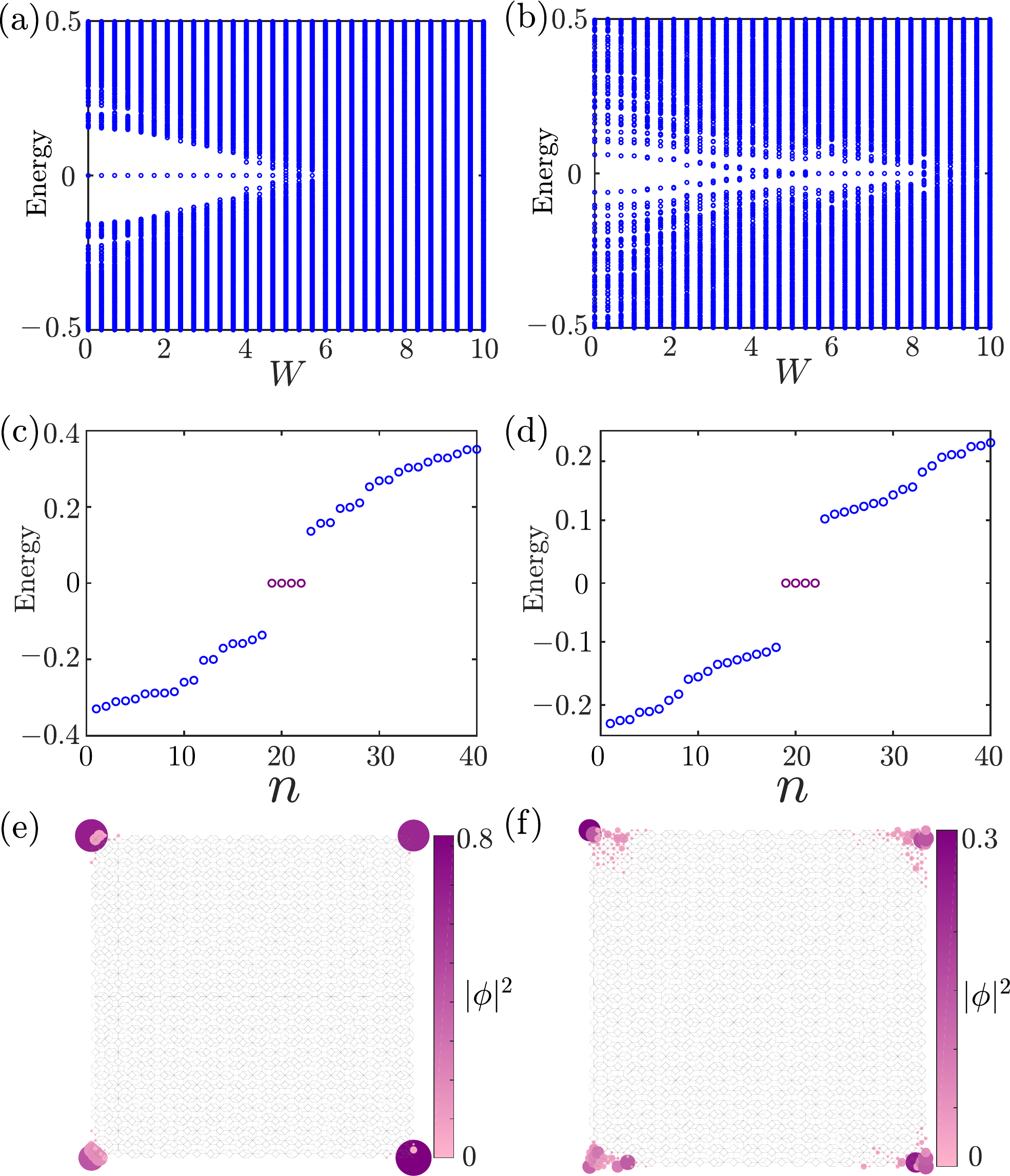} \caption{The eigenspectrum versus disorder strength $W$ with different initial states including: (a) a higher-order topological phase with $M=-1$ and (b) a topological trivial phase with $ M=1.6$. Some $200$ disorder configurations are performed with a square sample containing $4061$ sites. The energy modes near the zero energy for (c) a higher-order topological initial phase with $M=-1$, $W=1.5$ and (d) a trivial initial phase with $M=1.6$, $W=6.6$, respectively. (e) and (f) The wave-function distribution of the four in-gap states corresponds to (c) and (d), respectively. All calculations are based on the open boundary condition.}%
\label{fig4}
\end{figure}

In Figs.~\ref{fig4}(a) and \ref{fig4}(b), we plot the eigenspectrum of the open lattice as the function of disorders with different $M$. For the case of $M=-1$, the probability density of the four in-gap eigenstates near zero energy in the clean limit presents a picture with four corner states localized at the four corner of the lattice [see Fig.~\ref{fig7}(a) in the Appendix], indicating that the system hosts a SOTI phase. Upon introducing the disorder and increasing its strength, in Fig.~\ref{fig4}(a), it is shown that the midgap modes remain stable until $W\approx 5.5$, beyond which the bulk gap disappears, and the system is converted to an Anderson localized state phase. To further illustrate the stability of the SOTI phase, Figs.~\ref{fig4}(c) and \ref{fig4}(e) display the eigenspectrum and probability density of the in-gap eigenstates with $W=1.5$. It is found that the four corner states are stable under weak disorder, indicating that the second-order phase is robust against the weak disorder. For another case of $M=1.6$, the system is a normal insulator phase in the clean limit due to the fact that the middle four eigenstates near the zero energy are localized in the bulk [see Fig.~\ref{fig7}(b) in the Appendix]. With the increase in $W$, two topological phase transitions occur in Fig.~\ref{fig4}(b). First, in the region $4<W<8$, the four middle eigenvalues gradually tend to be degenerate near the zero energy, and midgap modes are generated, indicating that a phase transition from normal insulator phase to the HOTI phase occurs. Figures~\ref{fig4}(d) and \ref{fig4}(f) show the energy spectrum and probability density of the in-gap eigenstates of $H_{2}$ under the open boundary condition with $M=1.6$ and $W=6.6$. It is found that there are fourfold energy degenerate in-gap states under the condition of this set of parameters. What is more interesting is that the wave functions corresponding to these degenerate energies are all localized at the four corners of the lattice, which are the so-called corner states as shown in Fig.~\ref{fig4}(f). The corner states, induced by disorder, is strong evidence for the emergence of the HOTAI. Then, with the increase in the disorder strength, the higher-order phase converts to an Anderson insulator phase at $W\approx 8$ with the energy gap closure and all eigenstates being localized. Based on our calculations, we can draw two conclusions: first, the higher-order topological phase is relatively stable under weak disorder; second, disorder can also induce the higher-order topological phase in model II.

\begin{figure}[tp]
	\includegraphics[width=8.5cm]{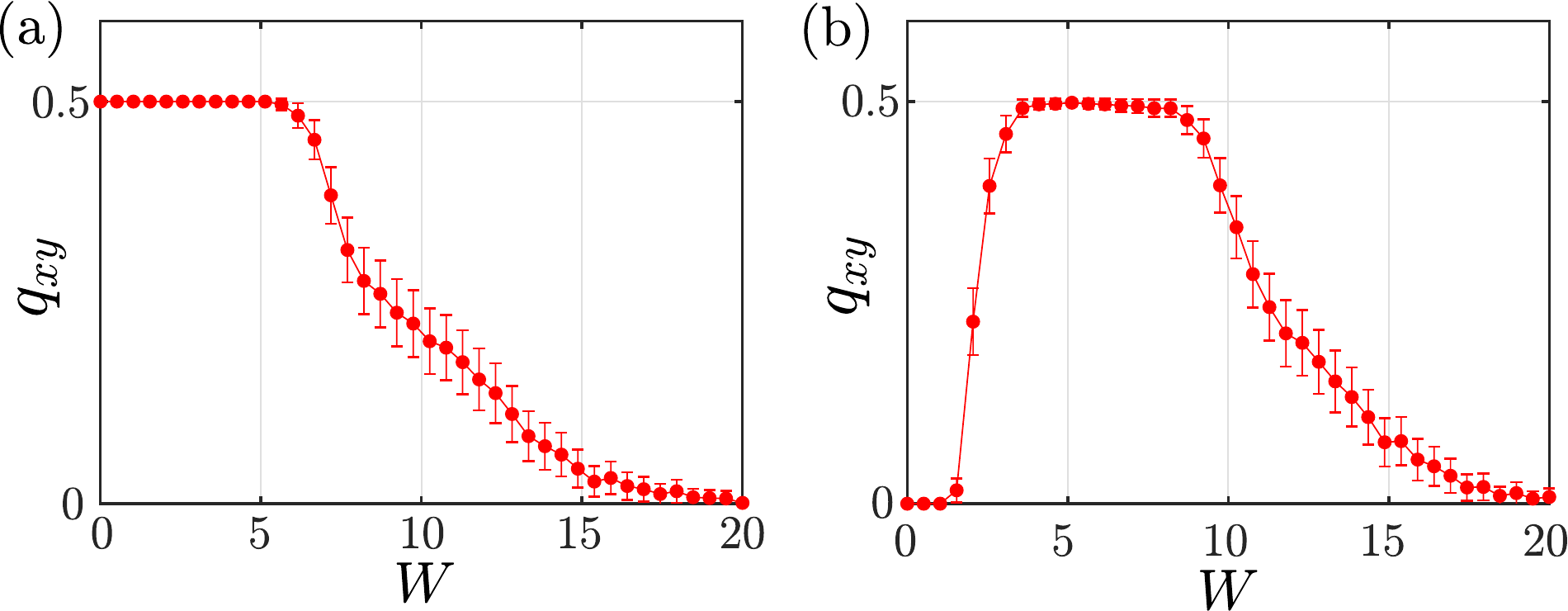} \caption{The real-space quadrupole moment $q_{xy}$ versus disorder strength $W$ with different initial states including (a) a higher-order topological phase with $M=-1$ and (b) a topological trivial phase with $ M=1.6$. The periodic boundary condition is taken, and $500$ disorder configurations are performed. The system is cut to a square sample which contains $1257$ sites.}

\label{fig5}
\end{figure}

As mentioned above, since both the chiral and the partial-hole symmetries are broken in model II with disorder, the necessary condition for the application of the real-space formula of the quadrupole moment is not be satisfied, and the real-space quadrupole moment should not be applied to characterize the higher-order topological phase in model II with the square sample. However, here we also try to calculate the real-space quadrupole moment $q_{xy}$ versus disorder strength with different $M$ as shown in Fig.~\ref{fig5}. Strikingly, it is found that the values of $q_{xy}$ are also quantized $1/2$ in certain disorder strength regions. For the case of $M=-1$ [see Fig.~\ref{fig5}(a)], the system hosts a SOTI phase with $q_{xy}=1/2$ in the clean limit, and a typical plateau is accompanied by quantized $q_{xy}$ until the strength of $W$ reaches a certain value ($W\approx 5.5$), indicating that the SOTI phase is robust against the weak disorder. However, the SOTI phase is eventually destroyed by strong disorder. For another case of $M=1.6$ [see Fig.~\ref{fig5}(b)], the system is a normal insulator phase with $q_{xy}=0$ in the clean limit. With the increase in $W$, two topological phase transitions occur, accompanied by $q_{xy}$ changing from $0$ to $0.5$ at $W\approx 4$ and returning to $q_{xy}=0$ at $W\approx 8$. In the region $4<W<8$, a remarkable plateau of quantized $q_{xy}=0.5$ appears, which indicates a SOTI phase induced by disorder. Thus, it is shown that the results given by $q_{xy}$ match well with the energy spectrum [comparing Figs.~\ref{fig4}(a) and \ref{fig4}(b) with Figs.~\ref{fig5}(a) and \ref{fig5}(b)]. It is implied that the validity of the operator-based formulation of the bulk quadruploe moment proposed by two pervious works \cite{PhysRevB.100.245134,PhysRevB.100.245135} is still an open issue. An intriguing question is whether this real-space quadrupole moment can still characterize the topology of the system in these situations without any symmetry constraint, and it will be further investigated in future work.

\subsection{Octagonal boundary condition}
\label{octagon}

\begin{figure}[t]
	\includegraphics[width=8.5cm]{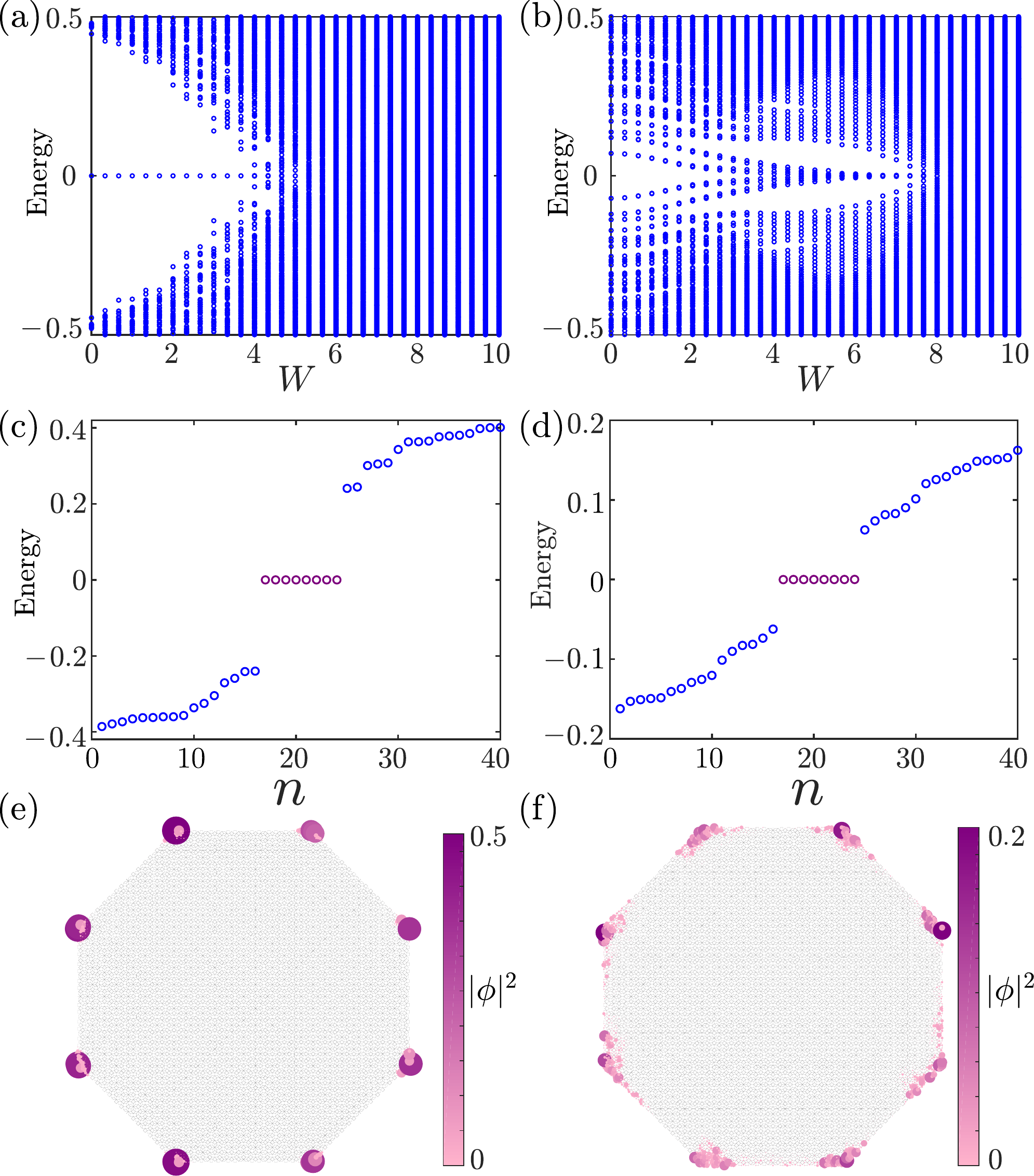} \caption {The eigenspectrum versus disorder strength $W$ with different initial states including (a) a higher-order topological phase with $M=-1$ and (b) a topological trivial phase with $ M=1.6$. Some $200$ disorder configurations are performed with a octagonal sample containing $13289$ sites. The energy modes near the zero energy for (c) a higher-order topological initial phase with $M=-1$, $W=2$ and (d) a trivial initial phase with $M=1.6$, $W=6.6$, respectively. (e) and (f) The wave-function distribution of the eight in-gap states corresponds to (c) and (d), respectively. All calculations are based on the open boundary condition.}%
\label{fig6}
\end{figure}

In Fig.~\ref{fig6}(a), we plot the eigenspectrum of the open lattice as the function of disorders with $M=-1$. The probability density of the eight in-gap eigenstates near zero energy in the clean limit presents a picture with eight corner states localized at the eight corners of the lattice [see Fig.~\ref{fig7}(c) in the Appendix], indicating that the system hosts a SOTI phase. Upon introducing the disorder and increasing its strength, the midgap modes remain stable until $W\approx 4$, beyond which the bulk gap disappears, and the system is converted to an Anderson localized state phase. To further illustrate the stability of the SOTI phase, Figs.~\ref{fig6}(c) and \ref{fig6}(e) display the eigenspectrum and probability density of the in-gap eigenstates with $W=2$. It is found that the eight corner states are stable under weak disorder.

In Fig.~\ref{fig6}(b), we plot the eigenspectrum of the open lattice as the function of disorders with $M=1.6$. The probability density of the middle eight eigenstates near the zero energy in the clean limit are localized in the bulk [see Fig.~\ref{fig7}(d) in the Appendix], indicating that the system hosts a trivial phase. Upon introducing the disorder and increasing its strength, a series of interesting changes occur in the energy spectrum. First, in the region $0<W<5.5$, the eight middle eigenvalues gradually tend to be degenerate near the zero energy, and midgap modes are generated, indicating that a phase transition from a normal insulator phase to a HOTI phase may occur. To verify this conclusion, we plot the eigenspectrum and probability density of the midgap eigenstates at $W=6.6$ as shown in Figs.~\ref{fig6}(d) and \ref{fig6}(f). It is shown that the eight midgap states are localized at the eight corners of the lattice, and these corner states are the powerful proof of disorder-induced HOTI. Then, with the increase in the disorder strength, the higher-order phase converts to an Anderson insulator phase at $W\approx 8$ with the energy gap closure and all eigenstates being localized.

\section{Conclusions and discussions}
In this paper, we investigate the disorder-induced higher-order topological phase transition in an Ammann-Beenker tiling quasicrystal. Two types of SOTI phases are considered: One is the quasicrystalline quadrupole insulator (model I), and the other is a quantum spin Hall insulator with a mass term which gapped the edge states and the topological corner states emerge (model II). Without disorder, model I (II) in the SOTI phase hosts gapless topological corner states protected by chiral ($C_{4}m_{z}$ or $C_{8}m_{z}$) symmetry and localized at the lattice corners. Based on calculating the quadrupole moment and the probability density of the middle gap eigenstates, it is found that in both models, the SOTI phases stay stable with weak disorder and are destroyed by strong disorder. More interesting is that the chiral symmetry-protected disorder-induced HOTAI is found by adding a certain strength of disorder to a topological trivial phase in model I. Meanwhile, a topological phase transition from a topological trivial phase to a HOTAI phase with topological corner states is also found in model II.

Based on the self-consistent Born approximation (SCBA), the disorder-induced topological phase transition forms a topological trivial phase to a topological nontrivial phase is attributed to the disorder that renormalizes the system parameters, such as the mass term, hopping term, and chemical potential \cite{PhysRevLett.103.196805,PhysRevB.95.094201,PhysRevLett.115.246603,PhysRevLett.116.066401,PhysRevB.95.245305,
PhysRevB.96.205304,PhysRevB.97.235109,PhysRevB.98.235159,PhysRevLett.125.166801}. However, the SCBA theory is invalid for the aperiodic systems, such as amorphous and quasicrystalline lattices which are the lack of translation symmetry. Up to now, there is not a well-defined theory to reveal the generating mechanism of TAI in aperiodic systems, and it is will be studied in the future work. Nevertheless, analogous to the generating mechanism of the TAI in a periodic system, we suppose that the generation of the TAI or HOTAI in the quasicrystalline system is also due to the renormalization of the parameters caused by disorder, and the initial trivial phase is converted to the HOTAI phase. In addition, disorder in model I does not destroy the chiral symmetry, and this symmetry also protects the topology of the system \cite{PhysRevLett.125.166801}.
In model II, the introducing of disorder has caused all of the symmetries of the system to be broken. It seems difficult to find a symmetry to guarantee the topology of the system, however, our calculations show that the HOTAI phases can be also induced by disorder in model II. We note that the quadrupole moment is easy to get a quantized value in model II. This may be caused by the following two points. One is that model I is more sensitive to the finite-size effect. Two is that the wave function of HOTI in model II is more local than model I. In previous work, Fu \emph{et al}. have proposed that, when the disorder is strong, the topological surface states exist, due to symmetries that are destroyed by disorder but remain unbroken on average \cite{PhysRevLett.109.246605}. Here, two key points are employed to guarantee that the averaged symmetries exist. One is that enough disorder configurations are needed for the average. The other one is the size of the systems should be large enough. Under these conditions, we suppose that the combined symmetries, such as $C_{4}m_{z}$ and $C_{8}m_{z}$ which are broken by random disorder will recovered statistically by taking an ensemble average \cite{PhysRevB.97.205110} and the HOTAI phases are protected by the average combined symmetries. More details about average symmetry which can keep the HOTI phases with disorder in quasicrystalline systems will be further investigated in the future work.

Recently, the HOTAI has been successfully implemented in a modified Haldane model based on electric circuits system \cite{PhysRevLett.126.146802}. Moreover, the quasicrystalline quadrupole topological insulatiors has been experimentally realized in electrical circuits \cite{lv2021realization}. Therefore, we propose an experimental setup to construct the quasicrystalline lattice in electronic circuits and realize the introduction of random disorder by changing the magnitude of the inductors and the capacitors. By this way, we believe that the HOTAI phase in the quasicrystalline system can be observed.
\label{Conclusion}

\section*{Acknowledgments}

B.Z. was supported by the NSFC (under Grant No. 12074107), and the program of outstanding young and middle-aged scientific and technological innovation team of colleges and universities in Hubei Province (under Grant No. T2020001). D.-H.X. was supported by the NSFC (under Grant No. 12074108). D.-H.X. also acknowledges financial support from the Chutian Scholars Program in Hubei Province.

\section*{Appendix: Wave function with square and octagonal boundary condition in the clean limit }

\begin{figure}[htb]
	\includegraphics[width=8.5cm]{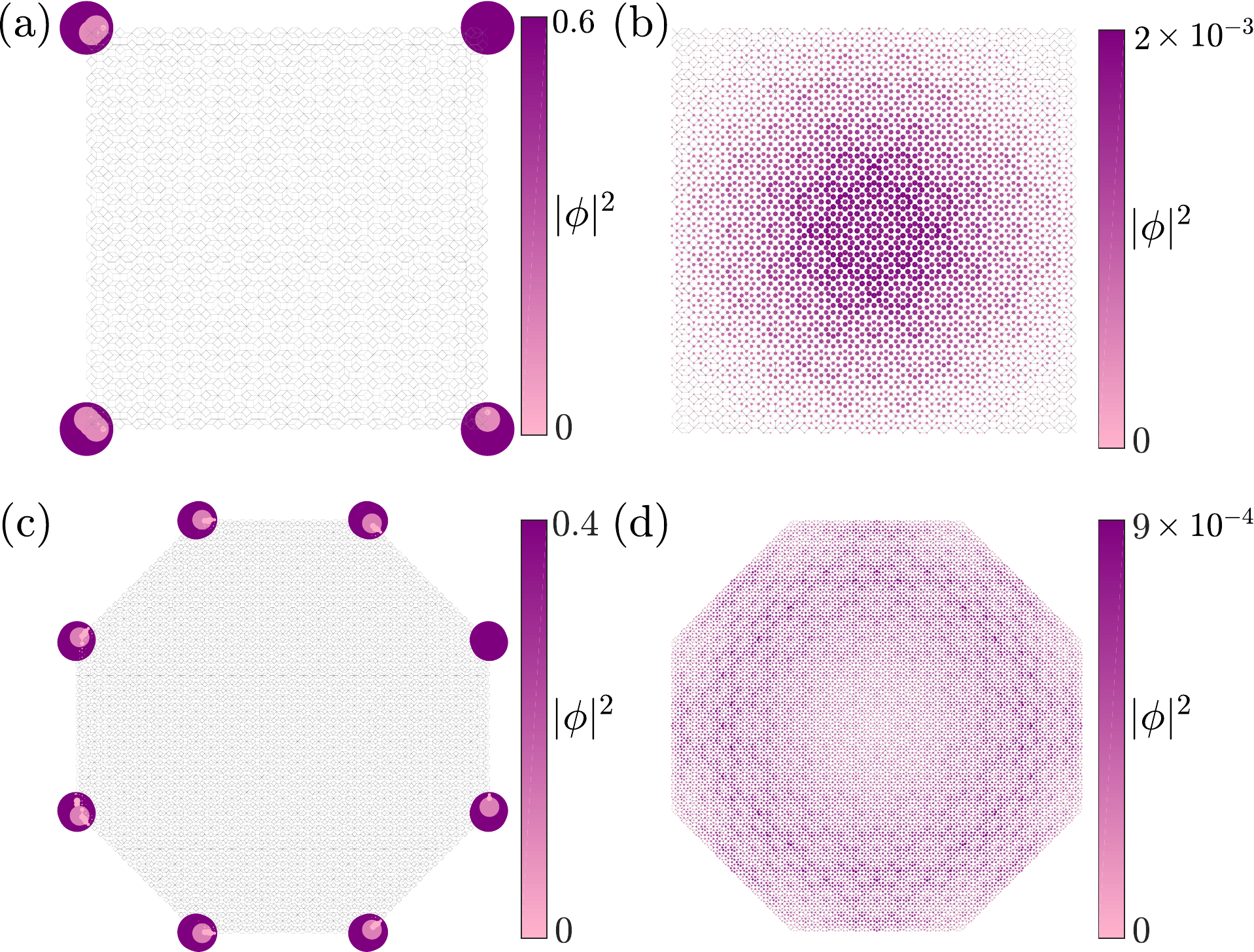} \caption{Probability density of the eigenstates in the clean limit with different Dirac mass (a) and (c)$M=-1$ and (b) and (d) $M=1.6$. The system is cut to a square sample which contains $4061$ sites for (a) and (b). For (c) and (d), the system is cut into a octagonal sample with $13289$ sites.}%
\label{fig7}
\end{figure}

In this Appendix, we plot the probability density of the four (eight) eigenstates which are nearest to zero energy in the clean limit with different Dirac mass $M$ to identify the initial phase of the system. All calculations are based on $H_{2}$. As shown in Figs.~\ref{fig7}(a) and ~\ref{fig7}(c), four and eight in-gap states symmetrically distributed at the corners of a quasicrystal octagon, indicating that the system is a HOTI phase at $M=-1$ in the clean limit. Meanwhile, when $M=1.6$, the system is in a topological trivial phase as shown in Fig.~\ref{fig7}(b) and ~\ref{fig7}(d).

\end{document}